\documentclass[epj]{svjour}
\usepackage{graphics}
\usepackage{color}
\newcommand{\eq}{{\,=\,}}

\def\La{\langle}
\def\Ra{\rangle}

\begin{document}
\title {Hydrodynamic Modeling and the QGP Shear Viscosity}
\subtitle{}
\author{Huichao Song\inst{1}
}                     
\institute{Department of Physics, The Ohio State University, Columbus, Ohio 43210, USA }
\date{Received: date / Revised version: date}
%
\abstract{In this article, we will briefly review the recent progress on hydrodynamic modeling
and the extraction of the quark-gluon plasma (QGP) specific shear viscosity with an emphasis on results
obtained from the hybrid model {\tt VISHNU} that couples
viscous hydrodynamics for the macroscopic expansion of
the QGP to the hadron cascade model for the microscopic
evolution of the late hadronic stage.
} 
\maketitle
\section{Introduction}
Heavy ion experiments at the Relativistic Heavy-
Ion Collider (RHIC) at Brookhaven National Laboratory
and the Large Hadron Collider (LHC) at CERN, have accumulated strong evidences for the creation of the quark gluon plasma (QGP)~\cite{Rev-Arsene:2004fa,Gyulassy:2004vg,Muller:2012zq}. The observation of strong collective flow and the successful descriptions from hydrodynamics demonstrate that the QGP is strongly coupled and behaves like an almost perfect liquid with a very small specific shear viscosity~\cite{Voloshin:2008dg,reviews}.  Using gauge/gravity (AdS/CFT) correspondence, Kovtun, Son and Starinets showed that there exists a lowest limit for the shear viscosity to entropy density ratio $\eta/s=1/4\pi$ (called as KSS bound) for a large class of strong-coupled quantum field systems (not including QCD)~\cite{Policastro:2001yc}. This raises the question how close to this limit is the specific shear viscosity of the QGP created at RHIC and the LHC.

Using weakly coupled QCD, one can calculate the QGP shear viscosity at very high temperatures from kinetic theory or from Kubo formula~\cite{Arnold:2000dr}. However, it is difficult to do a first principle calculation for the shear viscosity of the strongly coupled QGP created at RHIC and the LHC. It thus desirable to extract it from experimental data. Previous studies revealed that the anisotropic flow generated in relativistic heavy ion collisions is highly sensitive to the shear viscosity due to the rapid expansion of the QGP fireball, which leads to large viscous corrections from the shear velocity tensor~\cite{Teaney:2003kp}. Explicit viscous hydrodynamic simulations revealed that even the small specific shear viscosities at the KSS bound leads to significant suppression of elliptic and triangular flows~\cite{Romatschke:2007mq,Song:2007fn,Dusling:2007gi,Molnar:2008xj,Bozek:2009dw,Chaudhuri:2009hj,Qiu:2011hf,Schenke:2010rr}. In principle, this allows for an extraction of the specific QGP viscosity $\eta/s$ by tuning $\eta/s$ in viscous hydrodynamic calculations and fitting the results to the sensitive experimental observables. In practice, this procedure requires sophisticated theoretical modeling of the heavy-ion collisions, such as the initial conditions that include the fluctuation effects, an equation of state (EOS) that properly describes the speed of sound and  the QCD phase transition,  non-equilibrium kinetics and chemical composition in the late hadronic stage, etc.~\cite{Song:2008hj}.

For a realistic description of the late hadronic stage,
we developed the {\tt VISHNU} hybrid model that couples viscous hydrodynamics with a hadron cascade model which microscopically simulates the hadronic rescattering and the chemical and thermal freeze-out of varies hadron species through solving the Boltzmann equation with flavor dependent hadronic cross sections~\cite{Song:2010aq}. In this article, we briefly review the recent progress in hydrodynamic modeling and the extraction of the QGP viscosity from elliptic flow data with a special emphasis on the results obtained from the {\tt VISHNU} hybrid model~\cite{Song:2010mg,Song:2011hk,Song:2011qa,Heinz:2011kt}.\\

\section{Viscous hydrodynamics and the hybrid approach}
\subsection{Viscous hydrodynamics}
Viscous hydrodynamics is a macroscopic tool to describe the expansion of the QGP and subsequent hadronic matter. In this section, we briefly review the Israel-Stewart (I-S) viscous hydrodynamics in 2+1 dimension with the assumption of longitudinal boost invariance, which is solved by the {\tt VISH2+1} code developed at the Ohio State University around 2007~\cite{Song:2007fn}. For the work related to \"{O}ttinger and Grmela viscous hydrodynamics and recent progresses on 3+1-d viscous hydrodynamics (based on I-S formalism) without longitudinal boost invariance, please refer to Ref~\cite{Dusling:2007gi} and Ref.~\cite{Schenke:2010rr,Bozek:2011ua,Vredevoogd:2012ui} respectively.

{\tt VISH2+1} solves the equations for energy-momentum conservation and the 2nd order Israel-Stewart viscous equations~\cite{Israel:1976tn,Muronga:2001zk} (For simplicity, the net baryon number and heat conductivity are assumed to be zero).With Bjorken approximation~\cite{Bjorken:1982qr}, these equations can be conveniently written in the curvilear coordinates
$x^m=(\tau,x,y,\eta_s)$,  where $\tau=\sqrt{t^2{-}z^2}$
and $\eta_s=\frac{1}{2}\ln\bigl(\frac{t{+}z}{t{-}z}\bigr)$~\cite{Song:2007fn,Heinz:2005bw}:
\begin{eqnarray}
&&d_m T^{mn}=0, \qquad T^{mn} = e u^m u^n - p\Delta^{mn} + \pi^{mn},\ \ \
\label{eq2}\\
&&\Delta^{mr}\Delta^{ns} D\pi_{rs}=-\frac{1}{\tau_{\pi}}(\pi^{mn}{-}2\eta\sigma^{mn})
\nonumber\\ && \qquad \qquad \qquad \qquad
-\frac{1}{2}\pi^{mn} \frac{\eta T}{\tau_\pi}
       d_k\left(\frac{\tau_\pi}{\eta T}u^k\right),\\
        &&D \Pi
=-\frac{1}{\tau_{\Pi}}(\Pi+\zeta \theta)
 -\frac{1}{2}\Pi\frac{\zeta T}{\tau_\Pi}
       d_k\left(\frac{\tau_\Pi}{\zeta T}u^k\right).
       \label{I-S}
\end{eqnarray}

Here $e$ is the local energy density, $p$ is the local pressure, and $u^{m}$ is the flow 4-velocity. $\Pi$ is bulk pressure and $\pi^{m n}$ is the shear stress tensor.  $D\eq u^m d_m$ and $\nabla^m\eq\Delta^{ml}d_{l}$ ($\Delta^{mn} =g^{mn}{-}u^m u^n$) are the time and spacial derivative in the local comoving frame.
$\sigma^{mn}\eq\frac{1}{2}(\nabla^m u^n{+}\nabla^n u^m)-\frac{1}{3} \Delta^{mn} (d_k u^k)$ is the velocity shear tensor.

The shear viscosity $\eta$, bulk viscosity $\zeta$ and the
corresponding relaxation times $\tau_\pi$ and $\tau_\Pi$ are free parameters in viscous hydrodynamic calculations. The default settings in {\tt VISH2+1} are: $\eta/s = \mathrm{const.} $, $\tau_\pi= 3 \eta/(sT)$~\cite{Song:2007fn} and  $\zeta/s=0$ (We found that the bulk viscous effects are much smaller than the shear viscous effects due to the critical slowing down near phase transition~\cite{Song:2009rh}). For hydrodynamic calculations with temperature-dependent $\eta/s(T)$, please refer to~\cite{Shen:2011kn,Niemi:2011ix}

The equation of state (EOS) is an additional input for hydrodynamic simulations. The default EOS {\tt s95p-PCE} in {\tt VISH2+1} implements recent lattice results for the QGP phase and emphasizes the partially chemical equilibrium in the hadronic phase for temperatures below $T_{\mathrm{chem}} =165\ \mathrm{MeV}$~\cite{Huovinen:2009yb,Shen:2010uy}.

The initial entropy density profile for {\tt VISH2+1} are provided by two popular initial geometric models: Monte Carlo Glauber Model (MC-Glauber) and Monte Carlo KLN Model (MC-KLN)~\cite{Hirano:2009ah}. Due to the finite number of colliding nucleons, the initial eccentricities (which are the driving forces for the elliptic flow) fluctuate from event to event for a specific centrality bin. To account such fluctuating effects on average, we generate a large number of initial entropy density profiles from MC-Glauber or MC-KLN model, rotate each distribution either by aligning the participant plane or the reaction plane, and then average such rotated profiles to obtain one smoothed initial entropy density profile with participant plan eccentricity $\varepsilon_{part}$ or reaction plan eccentricity $\varepsilon_{rec}$. For calculation efficiency, early {\tt VISH2+1} and current {\tt VISHNU} perform calculations with such event-averaged initial conditions, which is called as \emph{single-shot simulations} (With $\varepsilon_{part}$ that is significantly larger than  $\varepsilon_{rec}$ in most central and most peripheral collisions, single-shot simulations with  smoothed initial profiles averaged in the participant plane partially accounts the fluctuation effects
imprinted in the flow data measured in the participant plane, such as $v_2\{2\}$).
The fluctuating profiles can also be directly put into {\tt VISH2+1} and {\tt VISHNU} in the \emph{event-by-event simulations}, resulting in fluctuating hadron spectra and flow that varies from event to event, which are then averaged to compare with the experimental data. Currently, event-by-event simulations have not been implemented in {\tt VISHNU}. For recent progress from event-by-event {\tt VISH2+1} simulations, one can refer to Ref.~\cite{Qiu:2011hf,Qiu:2011iv}.

The decoupling temperature that defines the hydrodynamic  freeze-out surface is generally set to $T_{\mathrm{dec}} = 100-120 \mathrm{MeV}$ to allow for sufficient evolution time to build up the phenomenologically required radial flow which controls the slopes of the hadron spectra and their dependence on hadron masses~\cite{Shen:2010uy}.

\subsection{{\tt \textbf{VISHNU}} hybrid model}

Although the implementation of the EOS {\tt s95p-PCE} properly accounts for the chemical freeze out for various hadron species, the pure hydrodynamic approach ultimately fails in the late hadronic stage due to the dramatic increases of viscous corrections which invalidate the fluid dynamical approach that requires near equilibrium. For a more realistic description of the evolution and decoupling of the late hadronic stage, we developed the hybrid model {\tt VISHNU}~\cite{Song:2010aq} by combining viscous hydrodynamics for the QGP fluid expansion with the hadron cascade model for the kinetic evolution of the hadronic resonance gas at a switching temperature $T_{sw}$.

The viscous hydrodynamics implemented in {\tt VISHNU} is {\tt VISH2+1}, which has been briefly described in Sec.~2.1. The hadronic cascade model used there is {\tt UrQMD} (Ultra-relativistic Quantum Molecular Dynamics model)~\cite{Bass:1998ca}, which microscopically simulates the evolution of the hadron resonance gas through the coupled Boltzmann equations with flavor-dependent cross-sections. The connection between {\tt VISHNU} and {\tt UrQMD} is realized through a Monte-Carlo event generator called {\tt H2O} which converts hydrodynamic output into particles profiles for further {\tt UrQMD} propagation by sampling the Cooper-Frye phase-space distribution (including viscous corrections) on the decoupling surface~\cite{Song:2010aq}.
The partially hadronic chemical equilibrium is naturally  imprinted in {\tt UrQMD} by simulating the dynamics of the hadronic gas with elastic, semi-elastic and inelastic collisions. By describing the hadronic rescattering and freeze-out procedure  microscopically, {\tt VISHNU} improves purely hydrodynamic models and eliminates the additional adjustable parameters required for the transport and freeze-out characteristics of the hadron phase, making it possible for a reliable extraction of the QGP viscosity from experimental data.

The default switching temperature $T_{sw}$ to switch hydrodynamics to the hadron cascade simulation is $165 \texttt{ MeV}$, which is chosen from the chemical freeze-out temperature measured at RHIC~\cite{BraunMunzinger:2001ip} and is approximately close to the QCD phase transition temperature from Lattice QCD simulations~\cite{lattice}. This is almost the highest temperature to implement {\tt UrQMD} without partonic degrees of freedom. It is also the lowest possible temperature for hydrodynamic description without introducing additional parameters for {hadronic viscous effects (including both viscosities and relaxations times) and sequential chemical freeze-out (which is realized in hydrodynamics by partial chemical equilibrium EOS with effective chemical potentials for different hadron resonances)}\footnote{{The bulk viscous effects near the phase transition are neglected here for simplicity.}}. {With such default setting for {\tt VISHNU}, the evolution of the system is described as a hydrodynamic expansion of the viscous QGP fluid with a chemical equilibrium EOS, followed by microscopic evolution of the hadron resonance gas through {\tt UrQMD}, in which the sequential chemical and thermal freeze-out for varies hadron species are realized through the elastic, in-elastic and semi-elastic collision rates in the Boltzmann equations.}

In Ref.~\cite{Song:2010aq}. we investigated whether the microscopic hadron cascade approach can be replaced by the macroscopic hydrodynamic approach (with temperature dependent specific shear viscosity $\eta/s(T)$ and partially chemically equilibrated   EoS {\tt s95p}{\tt-PCE} as input) by  varying the switching temperature $T_{sw}$ in {\tt VISHNU} simulations.
We found that with a constant $\eta/s$ as input, the elliptic flow shows a strong $T_{sw}$-dependence. After extracting a temperature dependent effective hadronic shear viscosity $(\eta/s)^{eff}(T)$ from integrated $v_2$ data {(with an assumption of short relaxation time $\tau_\pi=6\eta/(sT)$ from kinetic theory)}, we found that pure viscous hydrodynamics with $(\eta/s)^{eff}(T)$  could nicely fit the  $p_T$-spectra and differential elliptic flow $v_2(p_T)$ for identified particles calculated from {\tt VISHNU}. However, the extracted effective hadronic shear viscosity  depends strongly on the pre-hydrodynamic history, particulary the chosen value of the QGP shear viscosity due to the very possibly large relaxation time of the hadronic matter. It therefore does not represent the intrinsic transport properties of the hadronic matter, but a parameter that reflects some memories of the QGP transport properties.  {An extraction of both the shear viscosity and relaxation time from UrQMD demands a huge amount of computing resources, which is beyond our current scope of investigation.} Therefore, there exists no switching window below $T_{ch}$ in {\tt VISHNU}, where viscous hydrodynamics can replace the hadron cascade~\cite{Song:2010aq}. {To maintain the predictive power of {\tt VISHNU},
$T_{sw}$ is to suggested to set at 165 MeV, with which the other left free parameters can be fixed from experimental data}.

\begin{figure*}[tbh]
\center
\resizebox{0.85\textwidth}{5.5cm}{%
  \includegraphics{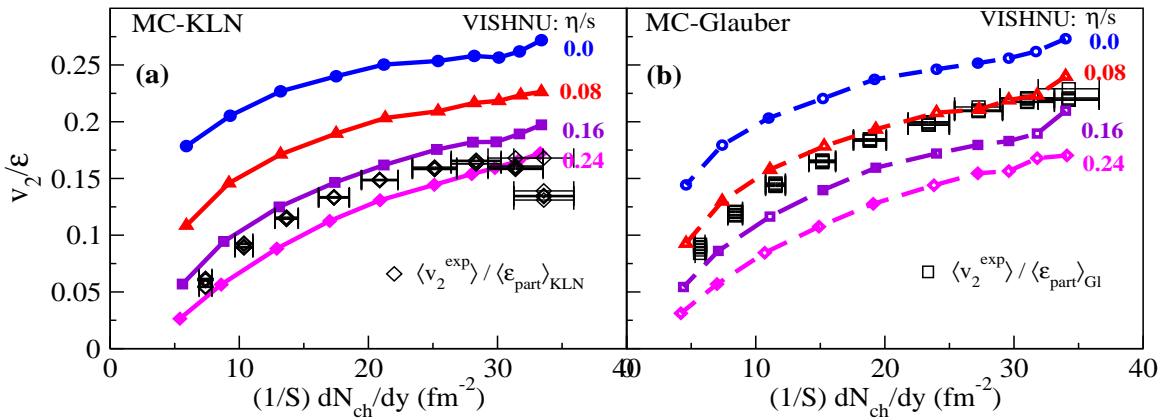}
}
\vspace*{0cm}  
\caption{(Color online) eccentricity-scaled elliptic flow as a function of final multiplicity per
area~\cite{Song:2010mg}}
\end{figure*}

\section{QGP viscosity from elliptic flow data: \\ \quad --the early attempt}

Elliptic flow and higher order flow coefficients  are important experimental observables for the bulk matter created in relativistic heavy ion collisions. In the language of hydrodynamics, pressure gradients convert the initial  deformations and inhomogeneities  of the fireball into fluid momentum anisotropies at different orders, which translates into the asymmetry of particle production as described by the flow coefficients. The shear viscosity controls the efficiency of this hydrodynamic conversion.  In the limit of zero shear viscosity, which corresponds to zero mean free path and instantaneous thermalization, the largest collective response is expected. The shear viscosity reduces the development of collective flow, leading to a suppression of the elliptic and triangular flow as observed by different groups~\cite{Teaney:2003kp,Romatschke:2007mq,Song:2007fn,Dusling:2007gi,Molnar:2008xj,Bozek:2009dw,Chaudhuri:2009hj,Drescher:2007cd,Lacey:2006pn,Xu:2007jv}.

Viscous hydrodynamics is a useful tool to study the viscous effects on the QGP fireball evolution and final observables.
During the past years, several groups have independently developed (2+1)-d~\cite{Romatschke:2007mq,Song:2007fn,Dusling:2007gi,Molnar:2008xj,Bozek:2009dw,Chaudhuri:2009hj} and (3+1)-d~\cite{Schenke:2010rr,Bozek:2011ua,Vredevoogd:2012ui} viscous hydrodynamic codes with/without longitudinal boost invariance for relativistic heavy-ion collisions at RHIC and LHC energies.  Past research showed that shear viscosity decelerates the longitudinal expansion, but accelerates the transverse expansion, leading to a shorter QGP lifetime, more radial flow and flatter hadron $p_T$-spectra~\cite{Song:2007fn}. More importantly, it was found that the elliptic $v_2$~\cite{Romatschke:2007mq,Song:2007fn,Dusling:2007gi,Molnar:2008xj,Bozek:2009dw,Chaudhuri:2009hj}  are very sensitive to the shear viscosity. Even the conjectured lower bound from the AdS/CFT correspondence $\eta/s=1/4\pi$ leads to a large suppression of  $v_2$. Thus one can extract the QGP shear viscosity from experimental data by a systematic fitting of $v_2$ as a function of collision energy, centrality, system size and etc.

The first attempt to extract the QGP viscosity from the elliptic flow data, using 2+1-d viscous hydrodynamics, was done by Luzum and Romatschke around 2008~\cite{Romatschke:2007mq}. They implemented two initial conditions from optical Glauber and KLN models and found that the $\sim 30\%$ uncertainties in initial eccentricity lead to $\sim 30\%$ uncertainties for the elliptic flow from viscous hydrodynamics with the same $\eta/s$, which then translate into $\sim 100\%$ uncertainties for the extracted value of the QGP shear viscosity. Two effects that are neglected in this work are {the off-equilibrium kinetics (or so-called highly viscous hadronic effects)~\cite{Hirano:2005xf,Demir:2008tr} }and the partially chemically equilibrated  nature~\cite{Teaney:2002aj,Kolb:2002ve} of the late hadronic evolution, which work against each other on influencing $v_2$ and may cancel to some extend. Furthermore, initial state fluctuations are neglected in their calculations and the effects from bulk viscosity was unclear around that time.  After making generous estimations for all these uncertainties, it appears that the averaged specific QGP shear viscosity, cannot exceed
the following conservative upper limit~\cite{Romatschke:2007mq,Song:2008hj}:
\begin{eqnarray*}
 \left.\frac{\eta}{s}\right|_\mathrm{QGP} < 5\times \frac{1}{4 \pi}.
\end{eqnarray*}

{\underline {A brief note on triangular flow:}}\\[-0.08in]

{ Recently, several groups extend single-shot hydrodynamic simulations to event-by-event ones, making it possible to investigate initial state fluctuations and higher order flow coefficients~\cite{Qiu:2011hf,Schenke:2010rr,Qiu:2011iv,Alver:2010dn,Petersen:2010cw,Holopainen:2010gz}.
It was found that triangular flow $v_3$  are also sensitive to the QGP shear viscosity as the elliptic flow $v_2$.
While the initial eccentricities $\varepsilon_2$ differ by O(20\%) between MC-KLN and MC-Glauber models,
the triangular deformation $\varepsilon_3$ are almost identical.  As a result, $v_3$ is much less sensitive to these two initializations compared with $v_2$. A systematic and combined analysis of  $v_2$ and $v_3$ together
may reduce the  initial conditions ambiguities and give an even accurate extracted value of the QGP shear viscosity~\footnote{
Currently, {\tt VISHNU} only concentrate on investigating $v_2$, since $v_3$ requires much more computing resources due to the event-by-event simulations.}.}

\section{QGP viscosity at RHIC and LHC energies: \\ \quad --the current status}

With the efforts from different groups, the elliptic flow is now widely accepted as a key observable to extract the QGP shear viscosity. However, it is also significantly affected
by the chemical composition and non-equilibrium kinetics of the late hadronic stage as well as the model uncertainties of the initialization eccentricity. With the newly developed viscous hydrodynamics + hadron cascade hybrid model {\tt VISHNU} on hand, which more realistically describes the hadronic stage and eliminates the related hadronic uncertainties, we made an extraction of the QGP shear viscosity from the corrected integrated $v_2$ data at top RHIC energies with emphasis on the remaining uncertainties related to initialization models and then extrapolated our calculation to LHC energies. Below is a brief summary of our recent results:

\subsection{QGP viscosity from RHIC integrated $v_2$ data}
{The hydrodynamic pressure gradients translate the initial fireball deformation $\varepsilon_x$ into fluid momentum anisotropy $\varepsilon_p$. Meanwhile, the shear viscosity suppresses the development of $\varepsilon_p$ during the fireball evolution through the anisotropic shear stress forces. The experimental observables that most directly related to $\varepsilon_p$ is the integrated elliptic flow $v_2^{ch}$ for all charged hadrons. While its distribution to the differential $v_2(p_T)$ for identical particles strongly depends on the chemical composition and radial flow of the hadronic matter, which constantly evolve during the fireball evolution even when $\varepsilon_p$ reaches saturation. Furthermore, $v_2(p_T)$  at higher $p_T$ region ($p_T > 1 \ \mathrm{GeV}$) is sensitive to the form of the non-equilibrium distribution function $\delta f$ and the inputting bulk viscosity. In contrast, such sensitivity is greatly reduced for the integrated $v_2$ (with $p_T$ spectra as a weighted function for the integration). We thus proposed to use the integrated $v_2^{ch}$ for all charged hadrons to extract the QGP shear viscosity~\cite{Song:2010mg,Song:2011hk}}.

In Ref.~\cite{Song:2010mg}, we found that the theoretical $v_2^{ch}/\varepsilon_x$ curves as a function of multiplicity density per overlap area $\mathrm{dN_{ch}/(dy S)}$  {are approximately universal, which are not very sensitive to the
initialization models and show clear separations between curves as the QGP specific shear viscosity increased by $1/4\pi$}. Furthermore, pre-equilibrium flow and bulk viscosity only slightly affect these theoretical curves, which are at or blow the order of 10\%. It thus preferable to extract the QGP viscosity from a comparison between the theoretical and experimental $v_2^{ch}/\varepsilon_x-\mathrm{dN_{ch}/(dy S)}$ curves, as shown in Fig. 1.  {the solid and dashed lines} with symbols  are the {\tt VISHNU} results with  different $(\eta/s)_{QGP}$ as input. Left and right panels correspond to  two different event-averaged,
smooth initial conditions, which are obtained through
averaging a large number of fluctuating initial entropy
densities (given by MC-Glauber or MC-KLN models) by aligning
the participant plane for each event. The experimental flow
measurements are generically contaminated by non-flow and
fluctuation effects to some extends, and are not suitable for direct comparison with these theoretical event-averaged $v_2$. Here we use the corrected elliptic flow data $\langle v_2^{exp} \rangle$ in the participant plane that removes non-flow and fluctuation effects, giving an almost universal curve for different corrected flow data from most central collision to most peripheral collisions~\cite{Ollitrault:2009ie}. While $\langle v_2^{exp} \rangle$ and $\mathrm{dN_{ch}/dy}$~\cite{:2008ez} are from experimental measurements, the theoretical inputs $\varepsilon$ and $S$  are calculated from the MC-Glauber and MC-KLN models. This leads to the differences in magnitude and slight changes in slope for the two experimental curves shown in Fig.1 left and right. As a result, the extracted value of $(\eta/s)_{QGP}$ from these two panels changes by a factor of 2 mainly due to the different $\varepsilon_x$ calculated from MC-KLN and MC-Glauber models. Recent event-by-event viscous hydrodynamic simulations showed that the triangular flow $v_3$ is also sensitive to the QGP shear viscosity~\cite{Schenke:2010rr,Qiu:2011hf} which suggests that, in the near future, the combined analysis of $v_2$ and $v_3$ from {\tt VISHNU} could yield an even more precise value of $(\eta/s)_{QGP}$ and may reduce the ambiguities in initial conditions from the hydrodynamic side. At this moment, we take the current uncertainties from the initial conditions and conclude from Fig.1 that {the averaged specific
shear viscosity of the QGP created at top RHIC energies is~\cite{Song:2010mg}:}
\begin{eqnarray*}
\frac{1}{4 \pi} <\left.\frac{\eta}{s}\right|_\mathrm{QGP} < 2.5\times \frac{1}{4 \pi}
\end{eqnarray*}

\begin{figure*}[tbh]
\center
\resizebox{0.7\textwidth}{!}{%
  \includegraphics{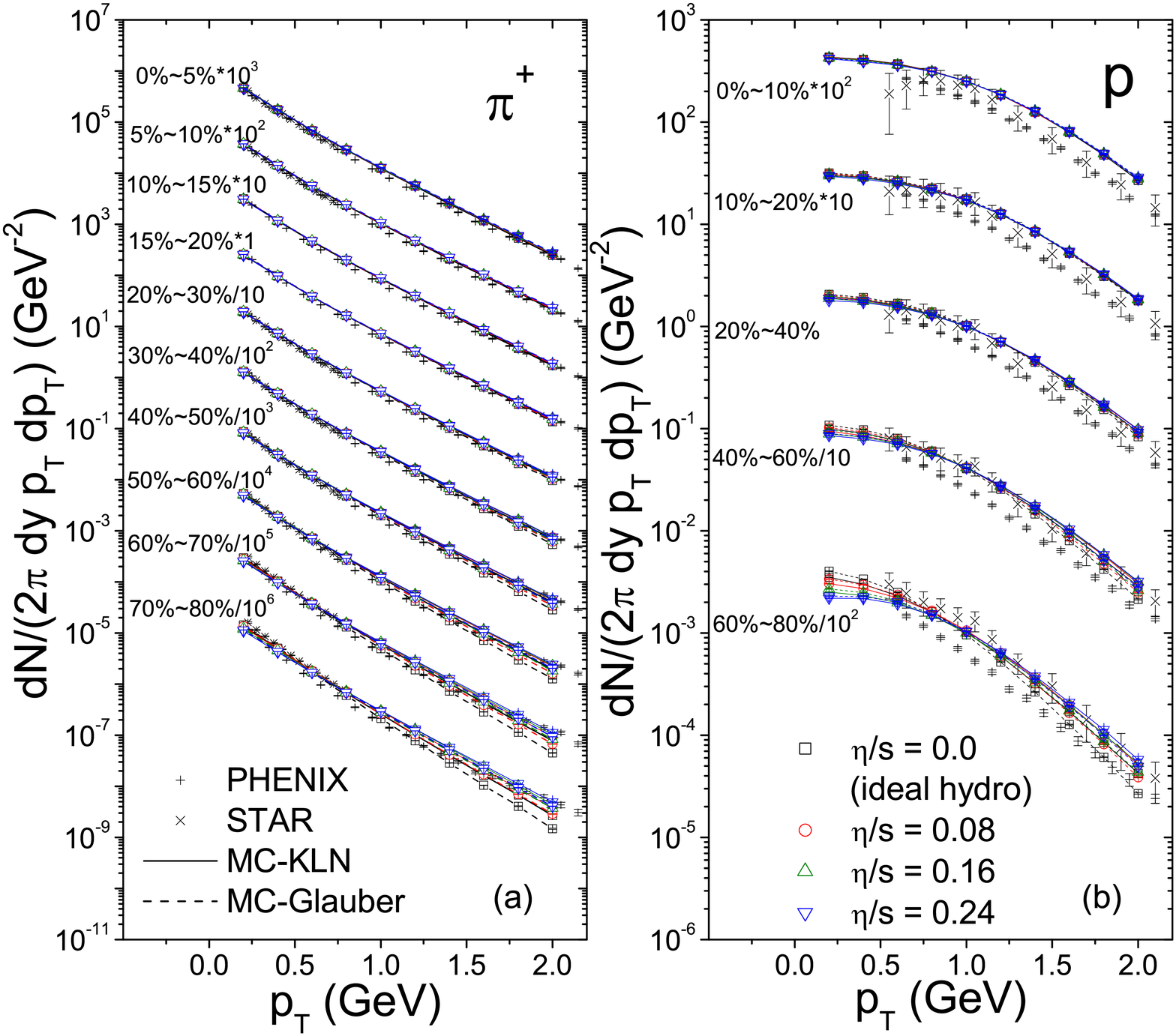}
}
\vspace*{0cm}       
\caption{(Color online) Transverse momentum spectra of pions and protons for 200 A GeV Au+Au Collisions at different centralities~\cite{Song:2011hk}.}
\end{figure*}

\subsection{Estimation of other effects}

In this section, we briefly estimate the residual effects on the extracted value of the QGP shear viscosity that are not directly included in the {\tt VISHNU} calculations shown in Sec 4.1 and Ref.~\cite{Song:2010mg}.\\[-0.08in]

\underline{Bulk viscosity:}\\[-0.08in]

Bulk viscosity also suppresses elliptic flow $v_2$ like shear viscosity~\cite{Bozek:2009dw,Song:2009rh}.  Whereas shear viscosity directly suppresses the development of flow anisotropies, bulk viscosity suppresses the development of radial flow, which indirectly influences the elliptic flow through changing the slope of the $p_T$ spectra and the lifetime of the QGP fireball. It is generally believed that the bulk viscosity to entropy density ratio $\zeta/s$ reaches a peak near the QCD phase transition $T_c$, while the value of this peak is murky, for which weakly coupled QCD~\cite{Arnold:2006fz}, strongly coupled N=4 SYM theory~\cite{Buchel:2007mf} and lattice QCD~\cite{Meyer:2007dy} gives dramatically different predictions. In Ref~\cite{Song:2009rh}, we found that the temperature dependent bulk relaxation time that describes the critical slowing down near the QCD phase transition, greatly offsets the effects from the strong growth of $\zeta/s$ near $T_c$, greatly reducing the bulk viscous suppression of $v_2$ even when $(\zeta/s)_{max}$ is very large. For simplicity, we neglected bulk viscosity in the {\tt VISHNU} calculations and predicted that the extracted values of the QGP specific shear viscosity will not be largely contaminated by the bulk viscosity due to critical slowing down near $T_c$. Although the inclusion of bulk viscous effects could reduce the extracted $(\eta/s)_{QGP}$, the total effects will be less than 20\%~\cite{Song:2009rh}.\\[-0.08in]

\underline{Event-by-event vs single shot calculation:}\\[-0.08in]

Current {\tt VISHNU} calculations employed single-shot hydrodynamics with event-averaged, smooth initial conditions (from MC-Glauber or MC-KLN model) followed by thousands of UrQMD simulations to obtain enough statistics for the spectra and $v_2$. This is a computationally efficient way to investigate the fluctuation effects on elliptic flow, and reduces the computing time by a factor of more than 20 when compared with the event-by-event simulations. However, it also raised the question on how much uncertainties it brings to the extracted $(\eta/s)_{QGP}$ due to the lack of real e-b-e simulations. A detailed comparison between e-b-e and single shot simulations from the pure hydrodynamics showed that e-b-e simulations reduce $v_2$ by O$(<10\%)$ for the same initial eccentricity~\cite{Qiu:2011hf}. This indicates that future e-b-e {\tt VISHNU} simulations {will reduce the extracted value of $(\eta/s)_{QGP}$ from the elliptic flow data by O$(<30\%)$}.\\[-0.08in]

\underline{Initial flow:}\\[-0.08in]

{The radial and elliptic flow may develop during the early evolution of classical gluon field and in the pre-equilibrium partonic stage before thermalization.~\cite{Kolb:2002ve,Vredevoogd:2008id}}. In Ref~\cite{Song:2010mg}, we studied the initial flow effects by tuning the initial starting time of hydrodynamics with the constraint from fitting the experimental $p_T$ spectra and final multiplicities. It turns out that the integrated $v_2$ is maximally increased by O$(10\%)$ by the initial flow. This translates into an increase of extracted value of $(\eta/s)_{QGP}$ {by O$(<30\%)$ from this effects alone}.\\[-0.08in]

\underline{The form of the viscous correction $\delta f$:}\\[-0.08in]

The form of the viscous correction $\delta f$ to the equilibrium distribution is an assumption in viscous hydrodynamic calculations. Although it does not directly influence the evolution of the QGP fireball, it affects the final $p_T$ spectra and differential $v_2(p_T)$ during the freeze-out procedure through the modified Cooper-Frye formula, which also influences the particle profiles propagated into the succeeding {\tt UrQMD} simulation in the {\tt VISHNU} hybrid approach.
Ref.~\cite{Dusling:2009df} showed that the quadratic and linear ansatz of $\delta f$ with different assumptions for the relaxation time could lead to an obvious difference in the
$v_2(p_T)$ for $p_T > 1 \ \mathrm{GeV}$. However, the effects at low $p_T$ are small, and thus only slightly influence the integrated $v_2$ within an order of $5\%$. Therefore, we choose the integrated $v_2$ as a preferable obserables to extract the QGP shear viscosity~\cite{Song:2010mg}. \\

\underline{{Other initialization models:}}\\[-0.08in]

{The main uncertainties for the extracted $(\eta/s)_{QGP}$ are from the undetermined initial conditions from MC-Glauber and MC-KLN models used in current {\tt VISHNU} calculations. The fluctuating sources for both models are from the fluctuating position distributions of nucleons in the colliding nuclei. In Ref.~\cite{Schenke:2012wb} and \cite{Muller:2011bb}, the additional quantum fluctuations for color changes are investigated under the framework of Color Glass Condensate(CGC).  Combing the Classical Yang-Mill's approach for Glasma field with the the impact parameter dependent saturation model (called IP-Glama model), Ref.~\cite{Schenke:2012wb} gave an modified initial eccentricity $\varepsilon_2$ and $\varepsilon_3$ that mostly lie between the ones from MC-Glauber and MC-KLN (except for the most central and most peripheral region)\footnote{Ref.~\cite{Muller:2011bb} investigated the transverse correlations for energy density fluctuations within the framework of Color Glass Condensate, but did not further calculate the initial eccentricity, since this needs to construct a new Monte-Carlo initialization generator to produce the initial energy density profiles with correlated fluctuations, rather than un-correlated ones from commonly used  MC-initialization models. This is still under  investigation~\cite{Ulrich}.}. In Ref.~\cite{Dumitru:2012yr}, the fluctuations of the initial gluon production are investigated accounting to a negative binomial distribution within the $k_T$ factorization approach of CGC. It gives an initial eccentricity  $\varepsilon_2$ that is very close to traditional MC-KLN one with only geometry fluctuations, but obvious larger $\varepsilon_3 -\varepsilon_5$ than the traditional MC-KLN ones. Considering that our current $(\eta/s)_{QGP}$ is extracted from the elliptic flow data driven by $\varepsilon_2$, the further implementation of other initializations from Ref~\cite{Schenke:2012wb} and Ref~\cite{Dumitru:2012yr}  will not change the current error bound of $(\eta/s)_{QGP}$.  As discussed in Sec.3, further investigations of elliptic, triangular flow and higher order flow harmonics together may help us to distinguish which initialization is preferred by the flow data and may give an even accurate value of the extracted $(\eta/s)_{QGP}$.} \\

To summarize briefly, {we extracted the QGP specific shear viscosity from the integrated elliptic flow data using {\tt VISHNU} with MC-Glauber and MC-KLN initializations as input, and found $1/(4\pi)<(\eta/s)_{QGP}<2.5/(4\pi)$. The width of this range is dominated by uncertainties of initial eccentricities
from these two initializations.} Small bulk viscous effects and proper event-by-event hydrodynamical evolution of fluctuating initial conditions may slightly reduce the integrated $v_2$. while early flow may slightly increase it.
Although they should be studied in more quantitative detail,
we expect the total uncertainty band translated to the extracted value of QGP shear viscosity  may slightly shift after cancelations.

\subsection{$P_T$ spectra and differential flow for identified particles}

After extracting the QGP shear viscosity $(\eta/s)_{QGP}$ from the integrated $v_2$ data for all charged hadrons, it is important to ensure that {\tt VISHNU} with $(\eta/s)_{QGP}$ also nicely describes the $p_T$ spectra and differential elliptic flow $v_2(p_T)$ for identified hadrons.  This was archived and documented in Ref~\cite{Song:2011hk}.

Fig.~2 shows the $p_T$ spectra for pions and protons at 200 A GeV Au+Au Collisions. The experimental data from STAR and PHENIX collaborations are compared with the {\tt VISHNU} calculations with MC-Glauber or MC-KLN initial conditions and different QGP specific shear viscosity as input.
From most central to semi-peripheral collisions, the theoretical curves are insensitive to initial conditions and $(\eta/s)_{QGP}$ and yield an excellent description of the experimental data. For most peripheral collisions, {\tt VISHNU} with finite QGP shear viscosity give a better fit of the experimental data, while the ideal fluid treatment shows a slightly steeper spectra for both pions and protons due to the insufficient development of radial flow~\cite{Song:2011hk}.

\begin{figure*}[tbh]
\center
\resizebox{0.7\textwidth}{!}{%
  \includegraphics{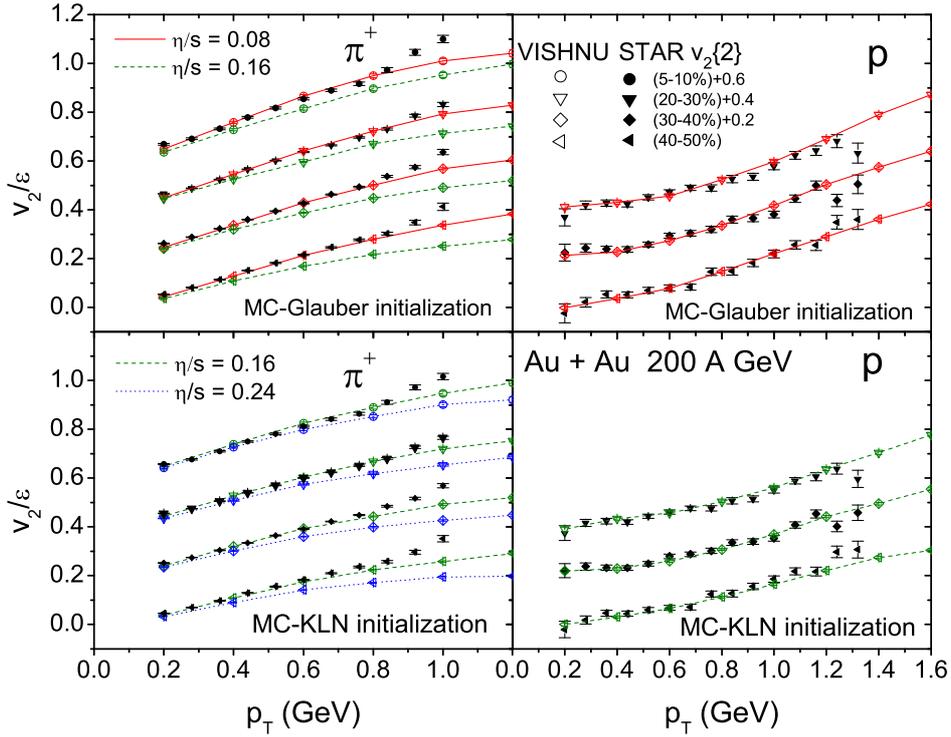} }
\vspace*{0cm}       
\caption{(Color online) eccentricity-scaled elliptic flow for  pions and protons at 200 A GeV Au+Au Collisions~\cite{Song:2011hk}. {Solid symbols denote reorganized experimental data $v_2\{2\}/\sqrt{\La \varepsilon_{part}^2}\Ra$, where the measurements $v_2\{2\}$ are from the STAR collaboration~\cite{Adams:2004bi}.  Solid and dashed lines with open symbols are theoretical $v_2/\varepsilon$ from
{\tt VISHNU}.}}
\end{figure*}

Fig.~3 shows the differential flow $v_2(p_T)$ for pions and protons from 200 A GeV Au+Au collisions at different centralities. Experimental data are from the STAR collaboration which is obtained from the 2-particle cumulant method, which measures $v_2\{2\}{\,\approx\,}\sqrt{\La v_2^2\Ra{+}\delta^2}$
with the contribution from event-by-event fluctuations $\sigma^2_{v_2}$ (where $\sigma^2_{v_2}=\La v_2\Ra^2-\La v_2^2\Ra$) and non-flow effects $\delta$~\cite{Ollitrault:2009ie,Alver:2008zza}. The theoretical curves are calculated from so-called ``one-shot" {\tt VISHNU} with the event-averaged initial conditions by aligning the participant plane which corresponds to a smooth initial entropy density profile with an eccentricity $\varepsilon_{part}$ that approximately equals the event averaged eccentricity $\La \varepsilon_{part}\Ra$. (The later is suppose to be the driving force for the event averaged elliptic flow $\sqrt{\La v_2^2\Ra}$).  Since such {\tt VISHNU} calculations are not the real event-by-event simulations, they can not be directly compared with the experimental $v_2\{2\}$ data. {We then compare the theoretical ratio $v_2/\varepsilon$ with the experimental ratio $v_2\{2\}/\sqrt{\La \varepsilon_{part}^2}\Ra$ by assuming that experimental $v_2\{2\}{\,\approx\,}\sqrt{\La v_2^2\Ra}{\,\approx\,}\frac{\La v_2\Ra}{\La
\varepsilon_\mathrm{part}\Ra}{\sqrt{\La \varepsilon_\mathrm{part}^2\Ra}}$ with a neglect of  the non-flow effects~\cite{Ollitrault:2009ie}}.

Fig.~3 demonstrates that with the $(\eta/s)_{QGP}$ extracted extracted from the $p_T$ integrated $v_2$ for all charged hadrons at 200 A GeV Au+Au collisions, {\tt VISHNU} yields an very nice description of the differential elliptic flow $v_2(p_T)$ for the identified hadrons (such as pions and protons) at different centrality bins~\cite{Song:2011hk}. The QGP shear viscosity read from Fig.~3, $(\eta/s)_{QGP}$ $\simeq (1/4\pi)$ for MC-Glauber and $(\eta/s)_{QGP}$ $\simeq (2/4\pi)$ for MC-KLN actually hit the lowest bound of $(\eta/s)_{QGP}$ from Fig.~1 for these two initial conditions respectively. The slightly lower $(\eta/s)_{QGP}$ here is due to non-flow effects in the experimental $v_2\{2\}$ data, which give a
positive contribution to the measured elliptic flow, leading to a slightly lower value of $(\eta/s)_{QGP}$ to fit the data.

\subsection{An extrapolation to the LHC energies}

The new measurement for 2.76 A TeV Pb+Pb collisions at the LHC shows a total charged hadron multiplicity density that is about a factor of 2.2 higher than the one for 200 A TeV Au+Au collisions at RHIC~\cite{Aamodt:2010pb,Aamodt:2010cz}, and indicates a $\sim30\%$ increase in the initial temperature of the QGP fireball. Meanwhile the ALICE collaboration at the LHC also discovered a $\sim 30\%$ increase in the integrated $v_2$ and a similar differential $v_2(p_T)$ when comparing with the one measured by STAR at top RHIC energies~\cite{Aamodt:2010pa}. This raises the question of how the QGP specific viscosity changes from RHIC to LHC or if one could extract a temperature dependent $(\eta/s)_{QGP}(T)$ from the currently available experimental data.

\begin{figure*}[tbh]
\vspace*{5mm}
\hspace*{0.2cm}
\resizebox{0.43\textwidth}{5.7cm}{
\includegraphics{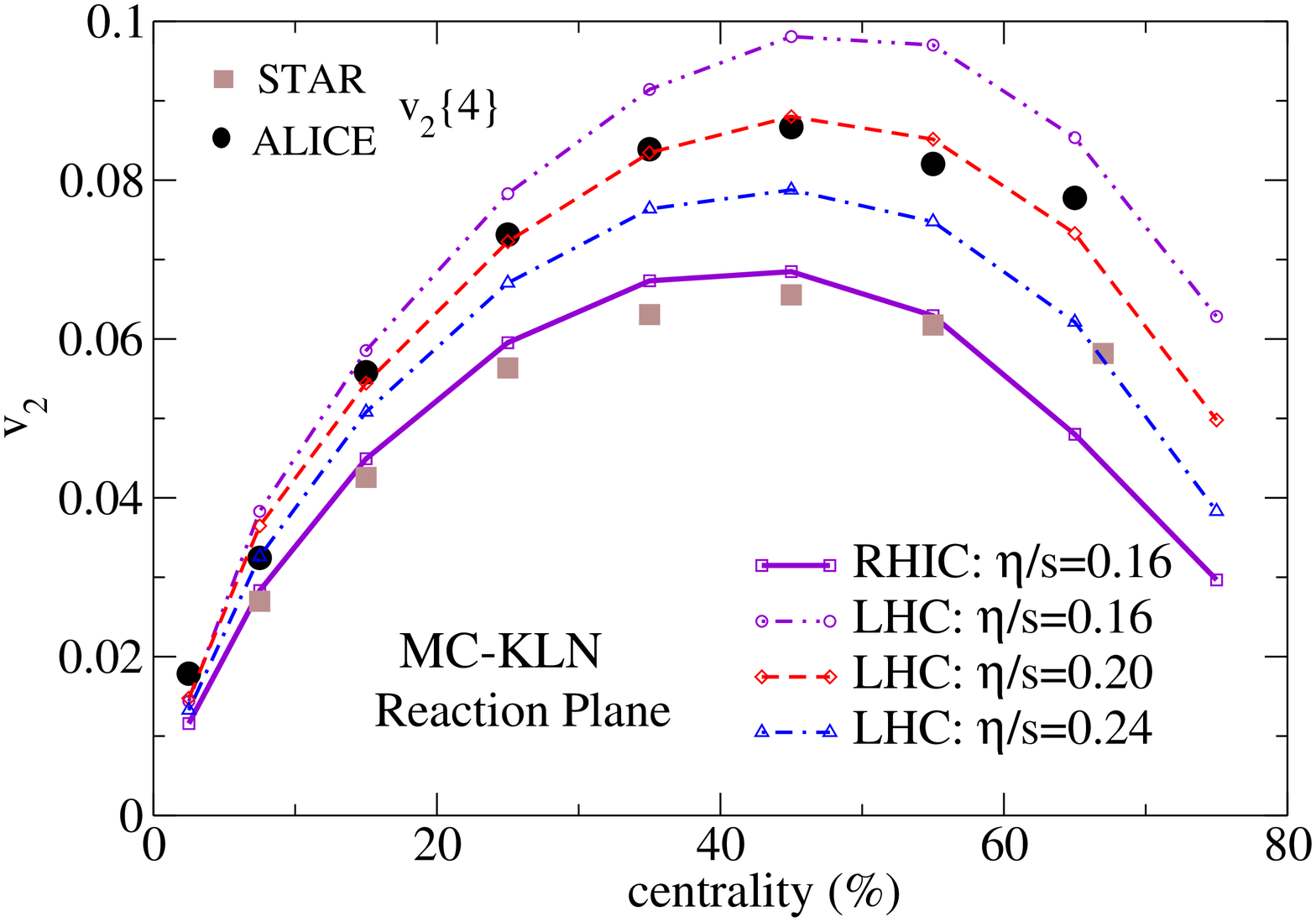}}
\hspace*{0.6cm}
\resizebox{0.48\textwidth}{5.6cm}{
\includegraphics{Fig5.eps} }
\vspace*{1mm}
\caption{(Color online) Integrated (left) and differential (right) $v_2$ for all charged hadrons at different
centralities at 2.76 A TeV Pb+Pb collisions and 200 A
GeV Au+Au collision~\cite{Song:2011qa}.}
\label{fig:1}       
\end{figure*}

Fig.~4 shows the comparison of the experimental and theoretical integrated and differential $v_2$ at 2.76 A TeV Pb+Pb collisions and 200 A GeV Au+Au collision~\cite{Song:2011qa}. The experimental $v_2$ data are from STAR and ALICE using a 4-particle cummulant method, which is supposed to measure $v_2$ in the reaction plane under the assumption of Gaussian fluctuations. The theoretical curves are from {\tt VISHNU} calculations with the event-averaged MC-KLN initial conditions created by aligning the reaction plane for each event. One finds a nice fit to the experimental data  with $(\eta/s)_{QGP} \sim 0.16$ at RHIC energies and $(\eta/s)_{QGP} \sim 0.20-0.24$ at LHC energies. Furthermore we also find a nice fit of the $v_2(p_T)$ for pion and protons at 2.76 A TeV Pb+Pb collisions  from {\tt VISHNU} with $(\eta/s)_{QGP} \sim 0.20$~\cite{Heinz:2011kt}.

The theoretical calculation shown in Fig.~4 is done by inputting a constant $(\eta/s)_{QGP}$, {which indicates that the \emph{averaged} specific shear viscosity (over the space time evolution of the QGP phase)} slightly increases with collision energy. However, this does not necessarily mean that QGP fluid is more viscous {at the higher temperature region reached by LHC since one needs to find one temperature-depended $(\eta/s)_{QGP}(T)$ that fits both the RHIC and LHC data rather than assuming two different constant $(\eta/s)_{QGP}$ for RHIC and LHC respectively}. A further study in Ref~\cite{Song:2011qa} shows that one can not uniquely constrain the forms of a temperature dependent $(\eta/s)(T)$ by fitting the spectra and elliptic flow at RHIC and LHC energies. Furthermore $v_2$ becomes more sensitive to the details of the stress tensor initialization at LHC energies~\cite{Shen:2011eg}. At the current stage, no firm conclusion can be drawn on whether or not the QGP fluid turns more viscous at the high temperatures probed by the LHC.

\section{Summary and Concluding Remarks}

In this article, we review recent results from the newly developed hybrid code {\tt VISHNU}~\cite{Song:2010aq} which combines the macroscopic viscous hydrodynamic description for the QGP fluid with the microscopic hadron cascade model for the subsequent evolution of the hadronic stage.  Using {\tt VISHNU} with an EOS that implements recent lattice results, we made an extraction of the averaged specific QGP shear viscosity from the integrated $v_2$ data in 200 A GeV Au+Au collisions that removes non-flow and fluctuation effects. We found that $1<4\pi(\eta/s)_{QGP}<2.5$ for the QGP created at RHIC, where the width of this range is mainly dominated by model uncertainties in the initial conditions~\cite{Song:2010mg}. Compared to the early extraction based on pure viscous hydrodynamics~\cite{Romatschke:2007mq}, this reduces the previous upper limit of $(\eta/s)_{QGP}$ by a factor of 2 due to the greatly improved description of the hadronic evolution. The $(\eta/s)_\mathrm{QGP}$ extracted from the centrality dependence of the integrated $v_2$ of all charged hadrons also provides consistent and nice description of the $p_T$-spectra and differential elliptic flow $v_2(p_T)$ for charged hadrons as well
as identified pions and protons over the entire range of collision centralities in 200 A GeV Au+Au collisions~\cite{Song:2011hk}. After extrapolating to the LHC energies, {\tt VISHNU} also yields a nice description of the integrated and differential $v_2$ for Pb+Pb collisions at 2.76 A GeV with roughly the same averaged QGP shear viscosity extracted at RHIC energies~\cite{Song:2011qa,Heinz:2011kt}.

All of these past {\tt VISHNU} results are done with single-shot hydrodynamic simulations with smooth initial conditions averaged over thousands of events with a rotation to align the "event plane"~\cite{Song:2010mg,Song:2011hk}. This is a computationally efficient way to include the effects of  fluctuations on the elliptic flow using the computationally demanding hybrid model {\tt VISHNU}. However, a detailed comparison between event-by-event and single-shot hydrodynamic simulations from {\tt VISH2+1} shows O(10\%) deviations for the elliptic and triangular flow, especially when the QGP viscosity approaches zero~\cite{Qiu:2011hf}.  Furthermore, higher order flow harmonics $v_4$, $v_5$ and $v_6$ cannot be realistically described by single-shot hydrodynamics through rotating and averaging the initial fluctuating profiles~\cite{Qiu:2011hf}. This indicates that more accurate extraction of the QGP viscosity from the flow data requires the full event-by-event simulations. E-b-e simulations from pure viscous hydrodynamic simulations revealed that the shear viscosity suppresses the triangular flow $v_3$ more than elliptic flow $v_2$~\cite{Qiu:2011hf,Schenke:2010rr}. This suggests that, in the near future, a combined analysis of $v_2$ and $v_3$ using the advanced hybrid model {\tt VISHNU} could yield a more precise extraction of the QGP shear viscosity and strongly reduce the uncertainties from the hydrodynamic initial conditions.

%

\end{document}